\documentclass[preprint,12pt]{elsarticle}
\usepackage{amssymb,amsmath,amscd}
\usepackage{amsfonts,bm,latexsym}

\journal{Physics Letters B}

\def\be{\begin{equation}}
\def\ee{\end{equation}}
\def\bea{\begin{eqnarray}}
\def\eea{\end{eqnarray}}

\def\nn{\nonumber}
\def\p{\partial}

\def\Star{\,^{\star}\!}

\begin{document}
\begin{frontmatter}

\title{General nonextremal rotating charged AdS black holes in five-dimensional
$U(1)^3$ gauged supergravity: A simple construction method}

\author{Shuang-Qing Wu} 
\address{College of Physics and Electric Information, China West Normal University,
Nanchong, Sichuan 637002, People's Republic of China}

\begin{abstract}
With the help of a generalized form of the metric ansatz found for the single-charge
case in a previous work [S.Q. Wu, Phys. Rev. D \textbf{83}, 121502(R) (2011)], I adopt
a simple algorithm to construct the most general nonextremal rotating charged black
hole solutions in five-dimensional $U(1)^3$ gauged supergravity. The general solution
that is interesting for testing the AdS$_5$/CFT$_4$ correspondence in M-theory, is
characterized by its mass, two unequal rotation parameters, three different $U(1)$ charges,
and a negative cosmological constant. The metric ansatz is very universal and illuminative,
it is not only especially suitable for constructing solutions with multiple different
electric charges in (un)gauged supergravities, but also for other dilatonic gravity theory.
\end{abstract}


\end{frontmatter}

\section{Introduction}

The discovery of the anti-de Sitter/conformal field theory correspondence has stimulated
a great deal of interest in the construction of rotating charged solutions in gauged
supergravities. Of particular interest are five-dimensional nonextremal rotating charged
anti-de Sitter (AdS) black holes within the maximal $SO(6)$-gauged $D = 5$, $\mathcal{N}
= 8$ supergravity, since this theory arises from the $S^5$ reduction of the type IIB
superstring. Black holes with Abelian gauge fields can carry three independent charges,
associated with the Cartan subgroup $U(1)^3$ of $SO(6)$. Equivalently, they can be viewed
as solutions in $D = 5$, $\mathcal{N} = 2$ gauged supergravity coupled to two additional
vector multiplets. The general black hole solution should be characterized by its mass,
two independent angular momenta, and three different charge parameters.

However, constructing such nonextremal rotating charged black hole solutions in $D = 5$
$U(1)^3$ gauged supergravity becomes quite a difficult problem, since unlike the case of
ungauged supergravity, there is no known solution-generating technique that could charge
up the already-known neutral rotating AdS$_5$ black hole solution found in \cite{HHT99}.
On the other hand, it is also quite different from the case of supersymmetric solutions
that receive considerable studies using the G-structure formalism which was introduced
first in $D = 5$ minimal gauged supergravity by \cite{GG03,GR04a} and then in  $\mathcal{N} = 2$
gauged five-dimensional supergravity in \cite{GR04b,GS05}. The basis for this technique
relies on solving the Killing spinor equation \cite{KS01b,LLPV07}. Nevertheless,
supersymmetric AdS$_5$ solutions can also have been successfully found as limits of
nonextremal solutions without using such a G-structure formalism.

Although there has been much progress over the last few years in obtaining new, nonextremal,
asymptotically AdS$_5$ black hole solutions of five-dimensional gauged supergravity theory,
one has not yet obtained a most general nonextremal solution with two rotation and three
charge parameters arbitrary. Apart from a known supersymmetric AdS$_5$ solution \cite{KLR06}
with two independent rotation and three unequal charge parameters, all the nonextremal black
hole families of exact rotating and charged solutions currently-found in $D = 5$ $U(1)^3$ gauged
supergravity theory have been obtained via applying two simplification strategies \cite{Chow10}:
either setting two angular momenta equal or setting certain charges equal. Since in these two
kinds of special cases, one could rely on an inspired ansatz to substantially minimize the
difficulty. So far, all the previously-obtained solutions for rotating charged AdS black
holes were not derived via a universal method other than via a combination of guesswork,
followed by explicit brute-force verifications of the field equations.

The currently-known nonextremal rotating charged black hole solutions in the five-dimensional
minimal and $U(1)^3$ gauged supergravities are as follows. Apart from a supersymmetric AdS$_5$
black hole solution \cite{KS01a}, a general nonextremal rotating charged black hole solution
\cite{CCLP05a} within $D = 5$ minimal gauged supergravity was obtained soon after the
discovery of its counterpart with two equal angular momenta \cite{CLP04a}. On the other hand,
for the most interesting case of $U(1)^3$ gauged supergravity, all the presently-obtained
nonextremal rotating charged AdS$_5$ black hole solutions are limited to the two special
cases: either with some charges equal, or with equal rotation parameters. In the much simpler
situation where two rotation parameters are set equal, the solution with three independent
charges was obtained in \cite{CLP04b}. For black holes with two independent rotation
parameters, a nonextremal solution where two charges are equal but the third one is set
to zero was found in \cite{CCLP05b}. This solution was then extended \cite{MP07} to the
case in which two of the three charges are set equal, with the third non-vanishing. In
addition, a solution with only one charge non-zero was constructed in \cite{CCLP07},
extending the special case given in \cite{CCLP05b} with only one rotation parameter.
The latter solution can be also viewed as that of Kaluza-Klein gauged supergravity in
which a different form for the single-charge solution was recently presented in \cite{SQWu11}.

Although these remarkable progresses have been made during the past years, the most general
nonextremal solution for a five-dimensional rotating AdS$_5$ black hole with two arbitrary
rotation parameters and three independent charges remains unknown until this work, since
no solution-generating technique in gauged supergravity is available for deriving the
nonextremal rotating charged AdS$_5$ black hole solutions from the neutral Kerr-AdS$_5$
solution \cite{HHT99}. Instead one has little option but to resort to brute-force
calculations, starting from a guessed ansatz in the special cases either with two
equal rotation parameters or with two or three charges equal to verify that all the
equations of motion are completely satisfied.

The main purpose of this work is two-fold. First, I propose a universal ansatz to construct
general nonextremal rotating charged AdS black hole solutions in gauged supergravities. The
metric ansatz essentially generalizes the one previously put forward in \cite{SQWu11} for
the single-charge case in all higher dimensions. It is especially convenient for constructing
black hole solutions with multiple different electric charges in gauged and ungauged
supergravity theories, but needless limited to such theories. Second, I construct a
general nonextremal solution for a rotating charged black hole in five-dimensional gauged
supergravity with two unequal angular momenta and with three independent charge parameters.
This finalizes the long-unsolved question of seeking the most interesting nonextremal AdS$_5$
black hole solution in five-dimensional $U(1)^3$ gauged supergravity. After that, the
conserved charges associated with the first law of thermodynamics are computed. Because
of the complexity of the solution, a more detailed version is expected to discuss other
interesting issues related to it.

The main purpose of this work is two-fold. First, I propose a universal ansatz to construct
general nonextremal rotating charged AdS black hole solutions in gauged supergravities. The
metric ansatz essentially generalizes the one previously put forward in \cite{SQWu11} for
the single-charge case in all higher dimensions. It is especially convenient for constructing
black hole solutions with multiple different electric charges in gauged and ungauged
supergravity theories, but needless limited to such theories. Second, I construct a
general nonextremal solution for a rotating charged black hole in five-dimensional gauged
supergravity with two unequal angular momenta and with three independent charge parameters.
This finalizes the long-unsolved question of seeking the most interesting nonextremal AdS$_5$
black hole solution in five-dimensional $U(1)^3$ gauged supergravity. After that, the
conserved charges associated with the first law of thermodynamics are computed. Because
of the complexity of the solution, a more detailed version is expected to discuss other
interesting issues related to it.

\section{Kerr-AdS$_5$ and three-charge
Cveti\v{c}-Youm solutions}

To construct the most general nonextremal rotating charged AdS$_5$ black hole solution in
$U(1)^3$ gauged supergravity, it is instructive to present the solution in an appropriate
form to which the already-known Kerr-AdS$_5$ vacuum solution \cite{HHT99} and the three-charge
Cveti\v{c}-Youm solution \cite{CY96a,CY96b} degenerate, respectively, in the uncharged case
and in the ungauged case. In terms of the Boyer-Lindquist coordinates, the Kerr-AdS$_5$ black
hole \cite{HHT99} in a non-rotating frame at infinity can be conveniently written as
\bea
ds^2 &=& -\frac{(1+g^2r^2)\Delta_{\theta}}{\chi_a\chi_b}\, dt^2
 +\Sigma\Big(\frac{r^2\, dr^2}{\tilde{\Delta}_r} +\frac{d\theta^2}{\Delta_{\theta}}\Big) \nn \\
&& +\frac{(r^2+a^2)\sin^2\theta}{\chi_a}\, d\phi^2
 +\frac{(r^2+b^2)\cos^2\theta}{\chi_b}\, d\psi^2 \nn \\
&&\quad +\frac{2m}{\Sigma}\Big(\frac{\Delta_{\theta}}{\chi_a\chi_b}\, dt
 -\frac{a\sin^2\theta}{\chi_a}\, d\phi -\frac{b\cos^2\theta}{\chi_b}\, d\psi\Big)^2 \, ,
 \label{KAdS5}
\eea
in which
\bea
&& \tilde{\Delta}_r = (r^2+a^2)(r^2+b^2)(1+g^2r^2) -2mr^2 \, , \nn \\
&& \Delta_{\theta} = 1 -g^2p^2 \, , \quad \Sigma = r^2 +p^2 \, , \quad
p^2 = a^2\cos^2\theta +b^2\sin^2\theta \, , \nn \\
&& \chi_a = 1 -g^2a^2 \, , \quad \chi_b = 1 -g^2b^2 \, . \nn
\eea
It is well-known that the above metric (\ref{KAdS5}) can be put in the Kerr-Schild form
after performing the coordinate transformations \cite{GLPP05}.

Now let us turn to the three-charge Cveti\v{c}-Youm solution \cite{CY96a,CY96b}. It is
remarkable to find that its metric can be expressed in the following suggestive form
\bea
ds^2 &=& (H_1H_2H_3)^{1/3}\bigg[ -dt^2
 +\frac{\Sigma r^2}{\bar{\Delta}_r}\, dr^2 +\Sigma\, d\theta^2
 +(r^2+a^2)\sin^2\theta\, d\phi^2 \nn \\
&& +(r^2+b^2)\cos^2\theta\, d\psi^2
 +\frac{2ms_1^2}{\Sigma H_1(s_1^2-s_2^2)(s_1^2-s_3^2)}k_1^2 \nn \\
&&\quad +\frac{2ms_2^2}{\Sigma H_2(s_2^2-s_1^2)(s_2^2-s_3^2)}k_2^2
 +\frac{2ms_3^2}{\Sigma H_3(s_3^2-s_1^2)(s_3^2-s_2^2)}k_3^2 \bigg] \, , \label{CY3c} \\
k_i &=& \frac{s_ic_1c_2c_3}{c_i}\Big(\frac{c_i^2}{c_1c_2c_3}dt -a\sin^2\theta\, d\phi
 -b\cos^2\theta\, d\psi\Big) \nn \\
&& +\frac{c_is_1s_2s_3}{s_i}\Big(b\sin^2\theta\, d\phi +a\cos^2\theta\, d\psi\Big) \, ,
\label{CY3ck}
\eea
where
\bea
\bar{\Delta}_r = (r^2+a^2)(r^2+b^2) -2mr^2 \, , \quad H_i = 1 +\frac{2ms_i^2}{\Sigma} \, . \nn
\eea
Three $U(1)$ Abelian gauge fields are given by $A_i = 2m/(\Sigma H_i)k_i$. Here and hereafter,
the short notations $c_i = \cosh\delta_i$ and $s_i = \sinh\delta_i$ ($i = 1, 2, 3$) are used.

It should be pointed out that the line element (\ref{CY3c}) is essentially a generalized form
of the metric ansatz previously-presented in \cite{SQWu11} for the single-charge case of the
Kaluza-Klein AdS black holes in all higher dimensions. To see it more apparently, one can make
coordinate transformations so that the metric (\ref{CY3c}) is re-expressed in terms of the
Kerr-Schild coordinates, see the Appendix for details. In doing so, I find the metric tensor
then can be written as
\bea
g_{\mu\nu} &=& (H_1H_2H_3)^{1/3}\Big[ \eta_{\mu\nu} +\frac{2ms_1^2}{\Sigma H_1(s_1^2
 -s_2^2)(s_1^2-s_3^2)}\bar{k}_{1\mu}\bar{k}_{1\nu} \nn \\
&& +\frac{2ms_2^2}{\Sigma H_2(s_2^2-s_1^2)(s_2^2-s_3^2)}\bar{k}_{2\mu}\bar{k}_{2\nu}
 +\frac{2ms_3^2}{\Sigma H_3(s_3^2-s_1^2)(s_3^2-s_2^2)}\bar{k}_{3\mu}\bar{k}_{3\nu} \Big] \, ,
\quad
\eea
and accordingly the three U(1) gauge potentials are $A_i = 2m/(\Sigma H_i)\bar{k}_i$ modulo
radial gauge transformations. In practice, one finds that it is more efficient to work with
the line element in terms of the Boyer-Lindquist coordinates rather than using the Kerr-Schild
coordinates since this will avoid doing some unnecessary and complicated coordinate
transformations.

Now if one of the three charges is set to zero, one obtains the two-charge Cveti\v{c}-Youm
solution in five dimensions. In arbitrary dimensions, the two-charge Cveti\v{c}-Youm black
hole solutions \cite{CY962c} can be recast into a similar form which is given in Appendix A
of \cite{Wu7d2c}. For cases with more charges, one can proceed in a similar manner. Therefore,
one reaches to a remarkable conclusion that all of already-known supergravity solutions with
multiple different electric charges in ungauged theories have a universal metric structure
that can be rewritten in the ansatz proposed in this way.

Having revealed the underlying metric structure shared by the black hole solutions in
ungauged supergravity theories, then what is the case of gauged supergravity solutions?
It is well-established that in the vacuum cases, one can simply replace the flat metric
$\eta_{\mu\nu}$ in the Kerr-Schild form by the AdS metric. Does this still work well in
the gauged supergravity cases? In a previous work \cite{SQWu11} that only dealt with the
single-charge case in Kaluza-Klein supergravity, it was demonstrated this indeed is the
case. What is more, it has been shown there \cite{SQWu11} that the same holds also true
for the cases of nonrotating and rotating charged AdS black holes with multiple pure
electric charges in gauged supergravity theories, including the recently-found
single-rotation solution \cite{Wu7d2c} in $D = 7$ dimensions and the $D = 4$ two-charge
solution \cite{Chow11a}. In particular, limited to the interesting $D = 5$ case, there
is no exception for the gauged supergravity solution with three independent charges
obtained in \cite{CLP04b} in the case where the two rotation parameters are set equal.

These remarkable facts demonstrate that all previously-known supergravity black hole
solutions with multiple different electric charges can be recast into a unified metric
ansatz, regardless they belong to ungauged theories or gauged ones. In other words,
supergravity black hole solutions in gauged theory inherit the same underlying metric
structure as their ungauged counterparts. This significant feature of supergravity
black hole solutions had not been exploited in any other previous work. Clearly, once
an exact solution in ungauged supergravity has been found and recast into the ansatz
given above, one can immediately replace the flat metric by the AdS metric to try the
corresponding gauged supergravity solution. [The remaining task is just to determine
the independent vectors $k_i$. However, if additional form fields with higher spin are
excited, the computation is still extraordinarily complicated.]

Therefore, it is suggested that the ansatz proposed here provides a universal method to
construct the most general rotating charged AdS black hole solutions with multiple pure
electric charges in gauged supergravity theories. This consists of one of the main results
of this work. Apart from the example given in \cite{Wu7d2c}, here I will provide one
more example that finalizes the goal for seeking the most interesting nonextremal AdS$_5$
black hole solution in five-dimensional $U(1)^3$ gauged supergravity. With the guidance
of the above ansatz, the explicit expression for the most general charged rotating AdS$_5$
solution with three unequal charges and with two independent rotation parameters has been
successfully found here using a simple construction procedure.

\section{General solution}

Before presenting the new exact solution, it is now in a position to first display the theory
of five-dimensional $\mathcal{N} = 2$ gauged supergravity coupled to two vector multiplets.
The Lagrangian for the bosonic sector is given by
\bea
\mathcal{L} &=& \sqrt{-g}\Big[R +4g^2(X_1 +X_2 +X_3) -3(\p\varphi_1)^2 -(\p\varphi_2)^2
 -\frac{1}{4}\sum_{i=1}^3X_i^2F_i^2\Big] \nn \\
&& +\frac{1}{4}\varepsilon^{\mu\nu\rho\sigma\lambda}F_{1\mu\nu}F_{2\rho\sigma}A_{3\lambda} \, ,
\eea
where $g$ is the gauge-coupling constant, and the quantities $X_i = H_i/(H_1H_2H_3)^{1/3}$
are formed from the two scalar fields $\varphi_1$ and $\varphi_2$  in the vector multiplets:
\be
X_1 = e^{\varphi_1 +\varphi_2} \, , \quad X_2 = e^{\varphi_1 -\varphi_2} \, , \quad
X_3 = e^{-2\varphi_1} \, .
\ee

To find the most general nonextremal solution in this theory, it is reasonable to first
assume that the metric of the new exact solution has the exquisite form
\bea
ds^2 &=& (H_1H_2H_3)^{1/3}\bigg[ -\frac{(1+g^2r^2)\Delta_{\theta}}{\chi_a\chi_b}\, dt^2
 +\Sigma\Big(\frac{r^2\, dr^2}{\Delta_r} +\frac{d\theta^2}{\Delta_{\theta}}\Big) \nn \\
&& +\frac{(r^2+a^2)\sin^2\theta}{\chi_a}\, d\phi^2
 +\frac{(r^2+b^2)\cos^2\theta}{\chi_b}\, d\psi^2 \nn \\
&&\quad +\frac{2ms_1^2}{\Sigma H_1(s_1^2-s_2^2)(s_1^2-s_3^2)}K_1^2
 +\frac{2ms_2^2}{\Sigma H_2(s_2^2-s_1^2)(s_2^2-s_3^2)}K_2^2 \nn \\
&&\qquad +\frac{2ms_3^2}{\Sigma H_3(s_3^2-s_1^2)(s_3^2-s_2^2)}K_3^2 \bigg] \, ,
\label{5d3cKAdS}
\eea
and the three $U(1)$ Abelian gauge potentials are given by $A_i = 2m/(\Sigma H_i)K_i$. This
is simply because the already-known equal-rotation ($a = b$) solution \cite{CLP04b} can be
written in such a form, without any doubt, there should be no exception for the general case
with two unequal rotations ($a \not= b$). Then the remaining thing is to determine three
unknown vectors $K_i$ and the radial unction $\Delta_r$ since all other expressions remain
unchanged. There are two different ways to arrive at this goal. One direct method is to
analytically solve all the equations of motion derived from the Lagrangian after substituting
the above metric ansatz and three gauge potentials into them. However, the subsequent
computation will be very involved. Instead, below I will present another simple algorithm
to implement this aim.

The construction procedure for the explicit solution consists of the following four steps:
(I) Write down the conjectured metric (\ref{5d3cKAdS}) using the ansatz proposed above; (II)
Construct three independent vectors $K_i$ using a lot of rules (see below); (III) Determine
the radial function $\Delta_r$; (IV) Verify that all the field equations are completely
satisfied. The first step has been completed in the above.

The most key point of this construction is how to write down the components of three vectors
$K_i$, which is summarized according to my previous experience as follows. (i) First, combining
the uncharged case (\ref{KAdS5}) and the ungauged case (\ref{CY3ck}), the expressions of three
$K_i$ are conjectured to behave like
\bea
K_i &\sim& C_{1i}\frac{s_ic_1c_2c_3}{c_i}\Big(\frac{c_i^2}{c_1c_2c_3}
 \frac{\Delta_{\theta}}{\chi_a\chi_b}\, dt -C_{2i}\frac{a\sin^2\theta}{\chi_a}\, d\phi
 -C_{3i}\frac{b\cos^2\theta}{\chi_b}\, d\psi\Big) \nn \\
&& -C_{4i}\frac{g^2ab\Delta_{\theta}}{\chi_a\chi_b}\, dt
 +\frac{c_is_1s_2s_3}{s_i}\Big(C_{5i}\frac{b\sin^2\theta}{\chi_a}\, d\phi
 +C_{6i}\frac{a\cos^2\theta}{\chi_b}\, d\psi \Big) \, , \label{3K}
\eea
since the expected new solution should include them as two special cases. The 18 undetermined
constants ($C_{1i}, \cdots, C_{6i}$) will be constructed in the next step. (ii) Second, using
the discrete inversion symmetry
\bea
&& a\to \frac{1}{ag^2} \, , \quad b\to \frac{b}{ag} \, , \quad m\to \frac{m}{a^4g^4} \, ,
\quad s_i\to ags_i \, , \nn \\
&& gt\to \phi \, , \quad \phi\to gt \, , \quad \psi\to \psi \, ,
\quad r\to \frac{r}{ag} \, , \quad p\to \frac{p}{ag} \, , \label{invsym}
\eea
and the interchange symmetry ($a\leftrightarrow b$ and $\phi\leftrightarrow \psi$) to write
down all 18 constants ($C_{1i}, \cdots, C_{6i}$). Noting that $c_i\to \sqrt{\Xi_{ia}}$ under
the inversion symmetry (\ref{invsym}), one can very easily obtain the explicit expressions
of all 15 undetermined constants except the three ones $C_{4i}$ since the fourth term in
the conjectured expressions (\ref{3K}) is new, it did not appear in the two special cases
mentioned before. However, there is no difficulty to determine them by using the above
symmetries at this step. The final expressions for three vectors $K_i$ are found to be
\bea
K_i &=& \frac{s_ic_1c_2c_3}{c_i}\frac{\sqrt{\Xi_{1a}\Xi_{2a}\Xi_{3a}
 \Xi_{1b}\Xi_{2b}\Xi_{3b}}}{\sqrt{\Xi_{ia}\Xi_{ib}}}\Big(\frac{c_i^2}{c_1c_2c_3}
 \frac{\Delta_{\theta}}{\chi_a\chi_b}\, dt \nn \\
&& -\frac{\Xi_{ia}}{\sqrt{\Xi_{1a}\Xi_{2a}\Xi_{3a}}}
 \frac{a\sin^2\theta}{\chi_a}\, d\phi -\frac{\Xi_{ib}}{\sqrt{\Xi_{1b}\Xi_{2b}\Xi_{3b}}}
 \frac{b\cos^2\theta}{\chi_b}\, d\psi\Big) \nn \\
&& +\frac{c_is_1s_2s_3}{s_i}\sqrt{\Xi_{ia}\Xi_{ib}}\Big(-\frac{c_1c_2c_3}{c_i^2}
 \frac{g^2ab\Delta_{\theta}}{\chi_a\chi_b}\, dt \nn \\
&& +\frac{\sqrt{\Xi_{1a}\Xi_{2a}\Xi_{3a}}}{\Xi_{ia}}
 \frac{b\sin^2\theta}{\chi_a}\, d\phi +\frac{\sqrt{\Xi_{1b}\Xi_{2b}\Xi_{3b}}}{\Xi_{ib}}
 \frac{a\cos^2\theta}{\chi_b}\, d\psi\Big) \, ,
\eea
in which $\Xi_{ia} = c_i^2 -s_i^2\chi_a$ and $\Xi_{ib} = c_i^2 -s_i^2\chi_b$.

At this time, it should be pointed out that the above construction of the vectors $K_i$ are
largely helped by the discovery of the two-charge solution \cite{Wu5d2c} in the theory and
the single-charge solution \cite{SQWu11} in the Kaluza-Klein theory. It is then realized
that the above-mentioned discrete inversion symmetry (\ref{invsym}) and the interchange
symmetry play a crucial role in the construction of the new exact solution. It should be
mentioned that the inversion symmetry (\ref{invsym}) was first observed in \cite{CLP07}
for the uncharged Kerr-AdS$_5$ solution, and then in the $D = 4$ rotating charged solutions
recently found by Chow \cite{Chow11a,Chow11b}. It should also note that the recently-found
single-charge solutions \cite{SQWu11} in all dimensions and the two-charge solution
\cite{Wu7d2c} in $D = 7$ dimensions exhibit the same discrete inversion symmetry. By
the construction procedure, the general solution presented here obviously endows with
the discrete inversion symmetry (\ref{invsym}).

Having given the full expressions of all three vectors $K_i$, it is of no difficulty to
obtain the explicit formulae of the radial function
\bea
\Delta_r &=& (r^2+a^2)(r^2+b^2)(1+g^2r^2) -2mr^2 +2mg^2\Big\{(s_1^2+s_2^2+s_3^2)r^4 \nn \\
&& -(s_1^2s_2^2 +s_1^2s_3^2 +s_2^2s_3^2)\big[(a^2+b^2 -2m)r^2 +a^2b^2(2 +g^2r^2)\big] \nn \\
&& +s_1^2s_2^2s_3^2\Big([(a+b)^2 -2m][(a-b)^2 -2m] -2g^2a^2b^2(2r^2+2m \nn \\
&& +a^2+b^2) +g^4a^4b^4\Big) +2mg^2a^2b^2\big[s_1^4s_2^4 +s_1^4s_3^4 +s_2^4s_3^4 \nn \\
&& -2s_1^2s_2^2s_3^2(s_1^2 +s_2^2 +s_3^2)\big] \Big\}\, ,
\eea
by requiring the metric determinant to be
\be
\sqrt{-g} = (H_1H_2H_3)^{1/3}\frac{\Sigma r\sin\theta\cos\theta}{\chi_a\chi_b} \, .
\ee

The final step is to mechanically verify that all of the field equations are indeed satisfied.
This point can be confirmed by the fact that the new exact solution naturally degenerates to
the uncharged case \cite{HHT99} and the ungauged case \cite{CY96a,CY96b}, it also reduces to
the  $D = 5$ single-charge case found in \cite{SQWu11} and the two-charge case \cite{Wu5d2c} in
their corresponding limits. In all of these cases, the two rotation parameters are independently
specified. When three $U(1)$ charges are independent of each other, the verification has also
been analytically implemented in the special case with just one rotation parameter and in the
case when the two angular momenta are set equal. However, limited to the memory of a personal
computer and the 32-byte Maple 7 program, the complete verification fails in the minimal
supergravity case (with three equal charges and with two unequal rotation parameters) and in
the most general case (with three different charges and with two unequal rotation parameters),
due to the complexity of the new exact solution. Nevertheless, the verification of the most
general solution can be very easily checked by choosing several groups of different numerical
values of ($m, g, a, b, \delta_1, \delta_2, \delta_3$).

The general solution constructed here is characterized by its mass, two unequal rotation
parameters, three different $U(1)$ charges, and a negative cosmological constant. It is
interesting for the AdS$_5$/CFT$_4$ correspondence in M-theory.

\section{Thermodynamics}

Having presented the explicit expression of the new exact solution, the last task of this work
is to examine its thermodynamics, leaving the other interesting issues to be further investigated
in a detailed version. The general black hole solution has a Killing horizon at $r = r_+$, the
largest positive root of $\Delta_r = 0$. On the outer horizon, the entropy and the Hawking
temperature are easily evaluated as
\be
S = \frac{\pi^2}{2\chi_a\chi_b r_+}\sqrt{W} \, , \qquad
T = \frac{\Delta^{\prime}_{r_+}}{4\pi\sqrt{W}} \, ,
\ee
where
\bea
W &=& \big[(r_+^2+a^2)(r_+^2+b^2) +2mr_+^2(s_1^2+s_2^2+s_3^2)\big]\Big\{(r_+^2+a^2)(r_+^2+b^2) \nn \\
&& +2mg^2[(a+b)^2 -g^2a^2b^2][(a-b)^2 -g^2a^2b^2]s_1^2s_2^2s_3^2 -4mg^2a^2b^2(s_1^2s_2^2 \nn \\
&& +s_1^2s_3^2+s_2^2s_3^2)\Big\} +8m^2r_+^2c_1c_2c_3s_1s_2s_3ab\sqrt{\Xi_{1a}\Xi_{2a}\Xi_{3a}\Xi_{1b}\Xi_{2b}\Xi_{3b}} \nn \\
&& +4m^2r_+^2\big[r_+^2+g^2a^2b^2(s_1^2+s_2^2+s_3^2)\big](s_1^2s_2^2+s_1^2s_3^2+s_2^2s_3^2) \nn \\
&& -4m^2\Big\{(a^2+b^2)(1+g^2a^2)(1+g^2b^2)r_+^2 +g^2\big[(a^4+b^4)r_+^2 \nn \\
&& +g^2a^4b^4(2+g^2r_+^2)\big](s_1^2+s_2^2+s_3^2) +g^2a^2b^2(a^2+b^2 \nn \\
&& +g^2a^2b^2)\big[2+g^2r_+^2(s_1^2s_2^2+s_1^2s_3^2 +s_2^2s_3^2)\big]\Big\}s_1^2s_2^2s_3^2 \nn \\
&& +4m^2g^4a^4b^4(s_1^4s_2^4+s_1^4s_3^4+s_2^4s_3^4) -8m^2r_+^2g^6a^4b^4s_1^4s_2^4s_3^4 \nn \\
&& +8m^3(r_+^2+g^2a^2b^2)s_1^2s_2^2s_3^2 \, . \nn
\eea
The angular velocities at the horizon are given by
\bea
\Omega_a &=& \frac{2mr_+^2}{W}\bigg\{ac_1c_2c_3\sqrt{\Xi_{1a}\Xi_{2a}\Xi_{3a}}\Big[r_+^2+b^2
 -2mg^2b^2\big(s_1^2s_2^2+s_2^2s_3^2 \nn \\
&& +s_1^2s_3^2 +2g^2b^2s_1^2s_2^2s_3^2\big)\Big]
 -bs_1s_2s_3\sqrt{\Xi_{1b}\Xi_{2b}\Xi_{3b}}\Big[(r_+^2+b^2)\chi_a^2 \nn \\
&& -2m\big[1+g^2a^2 +2g^2a^2(s_1^2+s_2^2+s_3^2) +g^2a^2(s_1^2s_2^2+s_2^2s_3^2 \nn \\
&& +s_1^2s_3^2)(1+g^2a^2) +2g^4a^4s_1^2s_2^2s_3^2\big]\Big]\bigg\} \, , \\
\Omega_b &=& \frac{2mr_+^2}{W}\bigg\{bc_1c_2c_3\sqrt{\Xi_{1b}\Xi_{2b}\Xi_{3b}}\Big[r_+^2+a^2
 -2mg^2a^2\big(s_1^2s_2^2+s_2^2s_3^2 \nn \\
&& +s_1^2s_3^2 +2g^2a^2s_1^2s_2^2s_3^2\big)\Big]
 -as_1s_2s_3\sqrt{\Xi_{1a}\Xi_{2a}\Xi_{3a}} \Big[(r_+^2+a^2)\chi_b^2 \nn \\
&& -2m\big[1+g^2b^2 +2g^2b^2(s_1^2+s_2^2+s_3^2) +g^2b^2(s_1^2s_2^2+s_2^2s_3^2 \nn \\
&& +s_1^2s_3^2)(1+g^2b^2) +2g^4b^4s_1^2s_2^2s_3^2\big]\Big]\bigg\} \, ,
\eea
while the three electrostatic potentials are computed as
\bea
\Phi_i &=& \frac{2mr_+^2}{W}\bigg\{\frac{c_is_i\sqrt{\Xi_{1a}\Xi_{2a}\Xi_{3a}\Xi_{1b}
 \Xi_{2b}\Xi_{3b}}}{\sqrt{\Xi_{ia}\Xi_{ib}}}\Big[(r_+^2+a^2)(r_+^2+b^2)
 +2m(s_1^2+s_2^2 \nn \\
&& +s_3^2-s_i^2)(r_+^2 -g^2a^2b^2s_i^2) -4mg^2a^2b^2s_1^2s_2^2s_3^2 -2m(a^2+b^2 \nn \\
&& -2m)\frac{s_1^2s_2^2s_3^2}{s_i^2}\Big] -\frac{abc_1c_2c_3s_1s_2s_3\sqrt{\Xi_{ia}\Xi_{ib}}}{c_is_i}
 \Big(g^2(r_+^2+a^2)(r_+^2+b^2) \nn \\
&& -2m(1+2s_i^2) +2mg^2(s_1^2+s_2^2+s_3^2-s_i^2)\big[r_+^2 -s_i^2(a^2+b^2)\big] \nn \\
&& +2mg^2\big[2m -(1+2s_i^2)g^2a^2b^2\big]\frac{s_1^2s_2^2s_3^2}{s_i^2}\Big)\bigg\} \, .
\eea

Clearly, the new exact solution is asymptotically AdS with the boundary metric
\bea
\lim_{r\to\infty}\frac{ds^2}{r^2} &=& -\frac{g^2\Delta_{\theta}}{\chi_a\chi_b}\, dt^2
 +\frac{dr^2}{g^2r^4} +\frac{d\theta^2}{\Delta_{\theta}}
 +\frac{\sin^2\theta}{\chi_a}\, d\phi^2 +\frac{\cos^2\theta}{\chi_b}\, d\psi^2 \, ,
\eea
with which one can choose two vectors
\bea
\hat{N}^a = \frac{\sqrt{\chi_a\chi_b}}{g\sqrt{\Delta_{\theta}}}(\p_t)^a \, , \qquad
\hat{n}^a = -g^2r^2(\p_r)^a \, , \nn
\eea
and use the procedure adopted in \cite{CLP06} to compute the conserved charges that obey
thermodynamical first laws.

Using the formulae
\bea
\mathcal{Q}[\xi] &=& \frac{1}{16\pi}\int_{S^3} d^3x \frac{\sin\theta\cos\theta}{g^3
 \sqrt{\chi_a\chi_b\Delta_{\theta}}} C_{acbd}\xi^{a}\hat{N}^b\hat{n}^c\hat{n}^d \nn \\
&=& \frac{\pi}{4}\int_0^{\pi/2} d\theta \frac{\sin\theta\cos\theta}{\Delta_{\theta}}
 r^4C_{artr}\xi^{a} \, ,
\eea
the conserved mass and the two angular momenta are computed as
\bea
M &=& -\mathcal{Q}[\p_t] = \frac{\pi m}{4\chi_a\chi_b}\bigg\{2c_1^2c_2^2c_3^2\Big(\frac{1}{\chi_a}
 +\frac{1}{\chi_b} -1\Big) +1 -\big(s_1^2s_2^2 +s_2^2s_3^2 \nn \\
&& +s_1^2s_3^2\big)(1 +\chi_a\chi_b)
 -2s_1^2s_2^2s_3^2\big[1 +(2 -\chi_a -\chi_b)\chi_a\chi_b\big] \bigg\} \, , \\
J_a &=& \mathcal{Q}[\p_{\phi}] = \frac{\pi m}{2\chi_a^2\chi_b}\big(ac_1c_2c_3\sqrt{\Xi_{1a}\Xi_{2a}\Xi_{3a}}
 -b\chi_a^2s_1s_2s_3\sqrt{\Xi_{1b}\Xi_{2b}\Xi_{3b}}\big) \, , \\
J_b &=& \mathcal{Q}[\p_{\psi}] = \frac{\pi m}{2\chi_a\chi_b^2}\big(bc_1c_2c_3\sqrt{\Xi_{1b}\Xi_{2b}\Xi_{3b}}
 -a\chi_b^2s_1s_2s_3\sqrt{\Xi_{1a}\Xi_{2a}\Xi_{3a}}\big) \, , \quad
\eea
while three electric charges can be evaluated as
\bea
Q_i &=& \frac{1}{16\pi}\int_{S^3} \Big(X_i^{-2}\Star F_i
 -\frac{1}{2}\epsilon_{ijk}A_j\wedge F_k\Big) \nn \\
&=& \frac{\pi m}{2\chi_a\chi_b}
 \Big(\frac{c_is_i\sqrt{\Xi_{1a}\Xi_{2a}\Xi_{3a}\Xi_{1b}\Xi_{2b}\Xi_{3b}}}{\sqrt{\Xi_{ia}\Xi_{ib}}}
 -g^2ab\frac{c_1c_2c_3s_1s_2s_3}{c_is_i}\sqrt{\Xi_{ia}\Xi_{ib}}\Big) \, . \qquad
\eea
These conserved charges are related by the first law of thermodynamics.

\section{Conclusions}

In this paper, I have proposed a universal ansatz that is not only especially suitable for
constructing black hole solutions with multiple unequal electric charges in gauged and
ungauged supergravity theories, but also for other dilatonic gravity theory. With the help
of the ansatz, a simple algorithm is then adopted to successfully construct the most general
nonextremal rotating charged AdS$_5$ black hole solution with two unequal angular momenta and
with three different charge parameters within five-dimensional $U(1)^3$ gauged supergravity.
I have also computed the conserved charges associated with the first law of thermodynamics.

Because of the complexity of the results presented in this article, many other interesting
issues leave to be further revealed. For example, a direct problem associated with the new
solution is to investigate the relations between the two previously-known solutions
\cite{CCLP05a,CLP04b} and the special cases where three charges are set equal or two
rotation parameters are set equal in the general solution presented here. On the other
hand, one would like to seek the supersymmetric limit of the general solution, just as
Ref. \cite{KLR06} did.

Furthermore, the metric ansatz proposed here is a most important development of the famous
Kerr-Schild ansatz, it needs further deeper investigations in different supergravity theories.
It is anticipated that the ansatz can open a new way towards constructing the most general
rotating charged black hole solutions with multiple pure electric charges yet to be found
in some other (un) gauged supergravity theories.

By the way, it is also interesting to ask whether the five-dimensional solutions with
non-spherical topology structure such as black string, black ring and black Saturn can be
recast into a similar ansatz. Another question is whether multi-black hole solutions and
multi-black ring solutions can be written in a similar formalism. Clearly, these points
deserve a deeper investigation since this will be greatly expand and deepen our knowledge
in the construction of exact solutions in higher dimensions.

\section*{Acknowledgements}

S.-Q. Wu is supported by the NSFC under Grant Nos. 10975058 and 10675051. The computation
within this work has been done by using the GRTensor-II program based on Maple 7. He is
grateful to Prof. Hong L\"{u} for useful discussions.

\section*{Appendix}

\renewcommand{\theequation}{A\arabic{equation}}
\setcounter{equation}{0}

In order to rewrite the Cveti\v{c}-Youm solution (\ref{CY3c})
\bea
ds^2 &=& (H_1H_2H_3)^{1/3}\bigg\{ \frac{\Sigma r^2}{\bar{\Delta}_r}\, dr^2 +\Sigma\, d\theta^2
 +(r^2+a^2)\sin^2\theta\, d\phi^2 \nn \\
&& +(r^2+b^2)\cos^2\theta\, d\psi^2 -\frac{\Sigma -2m}{\Sigma H_1H_2H_3}\Big[dt
 +\frac{2mc_1c_2c_3}{\Sigma -2m}\big(a\sin^2\theta\, d\phi \nn \\
&& +b\cos^2\theta\, d\psi \big) -\frac{2ms_1s_2s_3}{\Sigma}\big(b\sin^2\theta\, d\phi
 +a\cos^2\theta\, d\psi\big)\Big]^2 \nn \\
&& +\frac{2m}{\Sigma -2m}\big(a\sin^2\theta\, d\phi +b\cos^2\theta\, d\psi\big)^2
 \bigg\} \, ,
\eea
in terms of the Kerr-Schild coordinates as follows
\bea
ds^2 &=& (H_1H_2H_3)^{1/3}\bigg[ -d\bar{t}^2
 +\frac{\Sigma r^2}{(r^2+a^2)(r^2+b^2)}\, d\bar{r}^2 \nn \\
&& +\Sigma\, d\theta^2 +(r^2+a^2)\sin^2\theta\, d\bar{\phi}^2 \nn \\
&& +(r^2+b^2)\cos^2\theta\, d\bar{\psi}^2
 +\frac{2ms_1^2}{\Sigma H_1(s_1^2-s_2^2)(s_1^2-s_3^2)}\bar{k}_1^2 \nn \\
&& +\frac{2ms_2^2}{\Sigma H_2(s_2^2-s_1^2)(s_2^2-s_3^2)}\bar{k}_2^2
 +\frac{2ms_3^2}{\Sigma H_3(s_3^2-s_1^2)(s_3^2-s_2^2)}\bar{k}_3^2 \bigg] \, ,
\eea
in which
\bea
&& \bar{k}_i = \frac{s_ic_1c_2c_3}{c_i}\Big[\frac{c_i^2}{c_1c_2c_3}d\bar{t}
 -a\sin^2\theta\, d\bar{\phi} -b\cos^2\theta\, d\bar{\psi}
 +\frac{\Sigma r^2\, dr}{(r^2+a^2)(r^2+b^2)}\Big] \nn \\
&&\qquad +\frac{c_is_1s_2s_3}{s_i}\Big[b\sin^2\theta\, d\bar{\phi} +a\cos^2\theta\,
 d\bar{\psi} -\frac{2\epsilon mab\Sigma r^2\, dr}{(r^2+a^2)(r^2+b^2)\bar{\Delta}_r}\Big] \, , \nn
\eea
one can make the following coordinate transformations:
\bea
&& dt = d\bar{t} -\Big[c_1c_2c_3 +\frac{2(1-\epsilon)mab s_1s_2s_3}{(r^2+a^2)(r^2+b^2)}\Big]
 \frac{2mr^2}{\bar{\Delta}_r}\, dr \, , \nn \\
&& d\phi = d\bar{\phi} -\frac{2amr^2\, dr}{(r^2+a^2)\bar{\Delta}_r} \, , \quad
d\psi = d\bar{\psi} -\frac{2bmr^2\, dr}{(r^2+b^2)\bar{\Delta}_r} \, ,
\eea
where $\epsilon$ takes 0, or 1.

\end{document}